\documentclass[aps,pra,twocolumn,showpacs,showkeys,floatfix,superscriptaddress,10pt]{revtex4-1}
\usepackage[colorlinks,citecolor=blue,linkcolor=red,dvipdfm]{hyperref}
\usepackage{amsmath,amssymb,graphicx}
\usepackage{times}

\begin{document}

\title{Self-trapped spatially localized states in combined linear-nonlinear periodic potentials}

\author{Jincheng Shi}
\affiliation{State Key Laboratory of Transient Optics and Photonics, Xi'an
Institute of Optics and Precision Mechanics of CAS, Xi'an 710119, China}
\affiliation{University of Chinese Academy of Sciences, Beijing 100049, China}
\affiliation{Key Laboratory for Physical Electronics and Devices of the Ministry of Education \&
Shaanxi Key Lab of Information Photonic Technique, Xi'an Jiaotong University, Xi'an 710049, China}

\author{Jianhua Zeng}
\email{\underline{zengjh@opt.ac.cn}}
\affiliation{State Key Laboratory of Transient Optics and Photonics, Xi'an
Institute of Optics and Precision Mechanics of CAS, Xi'an 710119, China}
\affiliation{University of Chinese Academy of Sciences, Beijing 100084, China}

\begin{abstract}
We analyze the existence and stability of two kinds of self-trapped spatially localized gap modes, gap solitons and truncated nonlinear Bloch waves, in one-and two-dimensional optical or matter-wave media with self-focusing nonlinearity, supported by a combination of linear and nonlinear periodic lattice potentials. The former is found to be stable once placed inside a single well of the nonlinear lattice, it is unstable otherwise. Contrary to the case with constant self-focusing nonlinearity, where the latter solution is always unstable, here, we demonstrate that it nevertheless can be stabilized by the nonlinear lattice since the model under consideration combines the unique properties of both the linear and nonlinear lattices. The practical possibilities for experimental realization of the predicted solutions are also discussed.
\end{abstract}



\maketitle


\emph{Introduction}.---Periodic physical systems \cite{DEP,FPC,NW-PTSS}, where a finite band gap in the underlying linear spectrum always opens, have aroused a concern in the studies of wave localizations and dynamics properties, in past decades. These systems have been and still are being explored widely and actively in many branches of physics, such as atoms in crystalline lattices well known in solid-state physics \cite{SSP}, electromagnetic waves in photonic crystals \cite{DEP,FPC,PC,PCF} and optically induced photonic lattices in optics \cite{DFY,GS-PL,PL}, Bose-Einstein condensates (BECs) trapped in optical lattices \cite{GS-BEC-theory,LL-JPC, BEC-darkGap, DS-OL,SM-OL, ENP}.
The interaction of such periodic potentials and defocusing nonlinearity can reach a balance, leading to the formation of a novel localized mode---gap solitons---populating in band gaps of the underlying linear spectrum, challenging the well-established concept in homogeneous media where bright solitons do not exist in defocusing nonlinearity (they exist, however, only in focusing nonlinearity)  \cite{FPC, GS-PL, ENP, OSS}. Gap solitons have been realized experimentally in nonlinear optics (optical Bragg gratings \cite{GS-FBG}, waveguide arrays \cite{SS-AWA-2D, SGS-WA, GS-WA}, photonic lattices \cite{CD-GDS-HPL, GS-NOIL}), and recently in exciton-polariton BECs loaded in photonic lattices \cite{ GS-SM1, GS-SM2, GS-SM3}, as well as in a BEC trapped into an optical lattice \cite{GS-BEC}. The latter, in particular, is atom number limited [e.g.,бл250 atoms] \cite{GS-BEC}, making its observation inaccessible under ordinary experimental circumstances.

Recent studies have pushed the periodic physical systems to their nonlinear counterparts, namely, the systems with a nonlinear lattice \cite{NL1,NL2,NL3,NL4,NL5,NL6,NL,NW-BEC-PRMT,SD-INL,TD-NLOVOD,SS-NLHT,DSV-AL} ---the periodic potential acts merely on the nonlinear term (the coefficient in front of nonlinearity) wherein the strength and even the sign could be changed periodically. Physical mechanisms and properties of solitons and vortical ones (vortex solitons) supported by regular nonlinear lattices and a higher-order ones---competing cubic-quintic nonlinear lattices \cite{NL3,NL4}, as well as the combination of linear and nonlinear lattices \cite{LNLa, LNLb, LNL1, LNL2,LM-ABFM,DT-1D-OLII,SW-CLNPP,LPES-NSE} have recently been explored.

Besides the fundamental gap solitons, another novel type of self-trapped localized states that are being viewed as truncated nonlinear Bloch waves (TBWs) or gap waves \cite{GW, GW2, GW3, GW4, GW5,TBWS-OL}, which reside in the band gap of the linear Bloch-wave spectrum too, were also found in periodic systems. Keep in mind that the first BEC gap soliton \cite{GS-BEC} composes of few number of atoms, the TBWs, however, are not limited by atom numbers any more, namely, they can be generated for arbitrarily large initial atom numbers. Further  investigations demonstrated that the TBWs originate from the associated nonlinear Bloch waves, and are inherently linked to the gap solitons \cite{GW2, GW3}. The study of the TBWs dynamics in various nonlinear periodic systems is now being an important subject of scientific interest \cite{GW, GW2, GW3, GW4} and of practical implications \cite{GW5}.

The studies of gap solitons have so far mainly focused on defocusing (repulsive) nonlinearity \cite{ENP,NL}, while their properties under focusing (attractive) nonlinearity have not been well understood (only few papers focused on this topic) \cite{ GS-LI-2D, GS-fB, DGS-2D, S-NLL, GS-BEC-OL, GS-BEC-LS, BB-SW-2D, GS-MW, VS-MW}, and the experimental verification remains a blank. Still, although the stabilization of TBWs by the defocusing nonlinearity and linear lattice has been demonstrated both in theory \cite{GW2, GW3} and experiments \cite{GW, GW4, GW5}, its relevance in focusing nonlinearity has not yet been known, either. It is therefore a nontrivial issue of particular interest to survey them (localized states of gap types) systematically in focusing settings.

  In this article, we aim to find the self-trapped spatially localized wave structures of gap types, in the
forms of aforesaid gap solitons and TBWs, in one- and two-dimensional (1D and 2D) focusing nonlinear periodic system consisted of (linear) optical and nonlinear periodic lattice potentials---the combined linear-nonlinear lattices model. A noteworthy result is that the TBWs can be stabilized by the focusing segment of the nonlinear lattice, in contrast, they are always unstable at constant focusing nonlinearity. The stable gap solitons localized themselves within a unit cell of the nonlinear lattice, they are unstable otherwise. Stability regions of both solutions are obtained by linear-stability analysis and identified by direct simulations. We demonstrate that the localized solutions predicted here can be realized in experiments, e.g., by filling the photonic crystals with properly selected nonlinear materials in optics, and by integrating the optical lattices and Feshbach resonance managed by spatially patterned fields in BECs.

\emph{The model and its stability analysis}.---We consider dynamics of atomic BECs loaded into optical lattices with a periodic modulated Kerr cubic nonlinearity (nonlinear lattice) described by the mean-field generalized Gross-Pitaevskii or nonlinear Schr\"{o}dinger equation for the wave function $\psi (x,t)$:
\begin{equation}
i\psi _{t}=-\frac{1}{2}\nabla ^{2}\psi-V_L(r)\psi +V_{NL}(r)|\psi |^{2}\psi ,
\label{GPE}
\end{equation}%
where the optical lattice is expressed as $V_L(r)=V_0 \cos(\pi x)$ and $V_0 [\cos(2x)+\cos(2y)]$ in 1D and 2D cases (we fix $V_0=10$ throughout), and the nonlinear coefficient $V_{NL}(r)$ yields:
\begin{equation}
V_{NL}(x)=
\begin{array}{c}
g_0\cos(qx),~~
\end{array} \label{NL1d}
\end{equation}%
in 1D, and in 2D
\begin{equation}
V_{NL}(x,y)=
\begin{array}{c}
g_1+g_0[\cos(qx)+\cos(qy)].~~
\end{array} \label{NL2d}
\end{equation}%
Here the constant coefficients $g_0<0$ and $|g_1|<2 |g_0|$, to ensure the nonlinearity is self-focusing (thus $V_{NL}(r)<0$) at point $x=(y=)0$ at which the fundamental gap soliton will be placed. Replacing time $t$ by propagation distance $z$, the Eq. (\ref{GPE}) also describes electromagnetic wave propagation in optical nonlinear Kerr media.

\begin{figure}[tbp]
\begin{center}
\includegraphics[width=1.0\columnwidth]{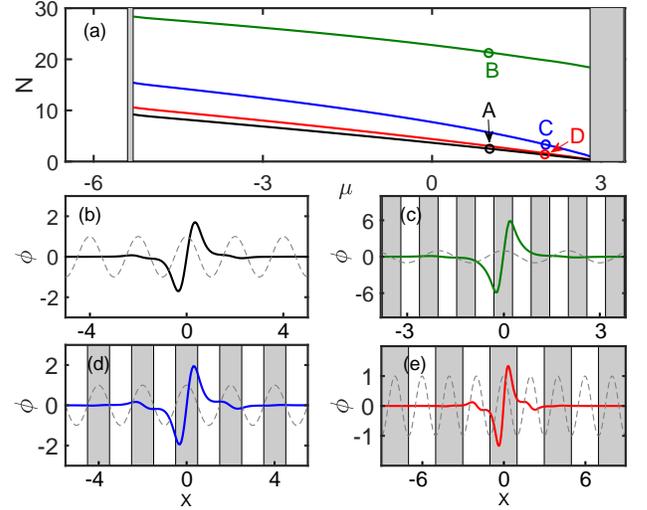}
\end{center}
\caption{(Color online)  (a) Norm $N$ versus chemical potential $\mu$ for family of 1D fundamental gap solitons within the first band gap at attractive nonlinearity ($g_0=-1$), in the cases of constant nonlinearity ($q=0$, black line), direct ($q=\pi$, blue line) and subharmonic ($q=\pi/2$, red line) commensurability as well as incommensurability ($q=\sqrt{3}\pi$, green line) between the linear and nonlinear periodic potentials. Typical stationary profiles of the fundamental gap solitons: at $\mu$=1 marked by circles A and B are, respectively, displayed in (b) and (c); at $\mu$=2 marked by circles C and D are displayed in (d) and (e). Shaded regions in (c, d, e), and in Fig. \ref{fig2}(c) below correspond to self-focusing segment of the nonlinear lattice. Dashed lines in (b, c, d, e) depict the profile of the linear lattice.}
\label{fig1}
\end{figure}

\begin{figure}[tbp]
\begin{center}
\includegraphics[width=1.0\columnwidth]{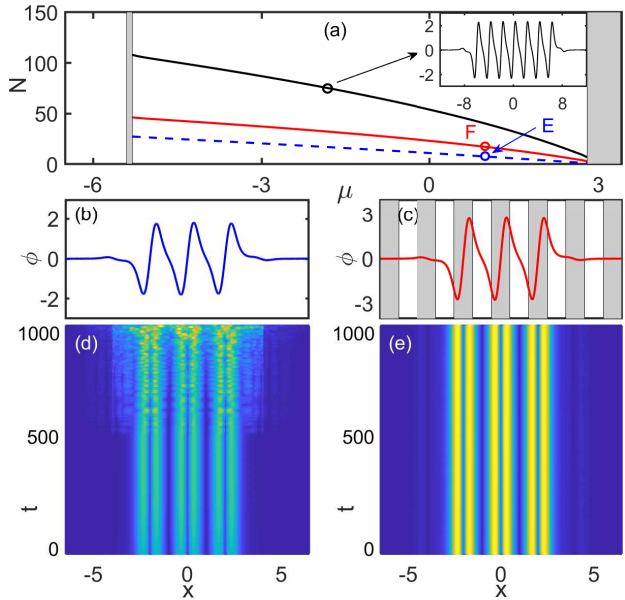}
\end{center}
\caption{(Color online) (a) Norm $N$ versus $\mu$ for families of 1D truncated nonlinear Bloch waves (TBWs) composed of six-peak (dashed-blue and red lines) and fourteen-peak (black) localized states under constant attractive nonlinearity (dashed-blue line) and nonlinear lattice (red and black lines). Typical profiles of the unstable (circle E) and stable (circle F) six-peak localized states under both nonlinearities are, respectively, depicted in (b) and (c); their perturbed evolutions are in (d) and (e).}
\label{fig2}
\end{figure}

We search stationary solution of wave function $\psi (x,t)=\phi (x)\exp(-i\mu t)$ at chemical potential $\mu$, then Eq. (\ref{GPE}) becomes
\begin{equation}
\mu\phi =-\frac{1}{2}\nabla ^{2}\phi+V_L(r)\phi +V_{NL}(r)\phi ^{3} .
\label{phi}
\end{equation}%

To tackle the stability problem of localized gap modes, the linear-stability analysis is adopted by taking the perturbed solutions as
\begin{equation}
\psi=[\phi + \upsilon \exp (\lambda t)+\omega^*\exp (\lambda^* t)]\exp (-i\mu t). \label{perturb}
\end{equation}%
Where $\phi$ is the so-found real stationary solution from Eq. (\ref{phi}) [$\phi$  is complex for vortex, the corresponding linear stability method also applies], $\upsilon$ and $\omega$ are, respectively, the real and imaginary parts of teeny perturbation modes, $\lambda$ is the corresponding growth rate or perturbation eigenvalue. Substituting expression (\ref{perturb}) into Eq. (\ref{GPE}), we get the eigenvalue problem for $\lambda$
\begin{equation}
i\lambda \upsilon=-\frac{1}{2}\nabla ^{2}\upsilon+[\mu+V_L(r)]\upsilon+V_{NL}(r)\phi ^{2}(2\upsilon+\omega),
\label{Eigena}
\end{equation}%
\begin{equation}
i\lambda \omega=-\frac{1}{2}\nabla ^{2}\omega+[\mu+V_L(r)]\omega+V_{NL}(r)\phi ^{2}(2\omega+\upsilon).
\label{Eigenb}
\end{equation}%
The solutions of perturbation equations (\ref{Eigena}) and (\ref{Eigenb}) give the stability condition of perturbed localized solution and its growth rate $\lambda$, from which the perturbed localized modes are always stable, provided that the real parts of all the eigenvalues obey $\lambda_r=0$.

The numerical results presented below are obtained in this procedure: the localized gap modes (in both 1D and 2D cases) are firstly constructed via stationary equation (\ref{phi}) by means of Newton's iteration, and then their stability is tested by eigenvalue problem relied on equations (\ref{Eigena}) and (\ref{Eigenb}) solving by a general finite-difference method, lastly the stability regions of the solutions thus found against small perturbations are given by direct simulations of equation (\ref{GPE}).

\emph{1D gap solitons and TBWs under focusing segment of the nonlinearity}.---We first study the stabilization of 1D gap solitons in the combined linear-nonlinear lattices model, with an emphasis on the role of self-focusing segment of the nonlinearity. In Fig. \ref{fig1} (a), the number of atoms (or norm) $N$ of the gap solitons within the first band gap is depicted as a function of chemical potential $\mu$, in cases of direct commensurability (spatial resonance) and subharmonic commensurability as well as incommensurability between the linear and nonlinear lattices for three values of $q$. Specifically, $q=\pi$ corresponds to commensurate lattices (both lattices with equal periods, $P_{lin}= P_{nonlin} $), $q=\pi/2$ represents the subharmonic commensurability [$P_{lin}= (1/2)P_{nonlin} $], and $q=\sqrt{3}\pi$ implies incommensurability. Meanwhile, for comparison, the dependence $N(\mu)$ under uniform nonlinearity ($q=0$) is also taken into account. Systematic simulations demonstrate that the entire gap soliton family under self-focusing segment of the cubic nonlinearity is stable at different values of $q$, obeying the well-known Vakhitov-Kolokolov criterion $dN/d\mu<0$ [see Fig. \ref{fig1} (a)] \cite{VK}, contrary to self-defocusing scenario where the gap solitons adhere to the inverted Vakhitov-Kolokolov criterion $dN/d\mu>0$ \cite{LNL1, LNL2}. Typical profiles of such gap solitons under incommensurability and the two kinds of commensurabilities are respectively illustrated in Figs. \ref{fig1} (c), \ref{fig1} (d) and \ref{fig1} (e). It is observed that the gap soliton is in the form of a dipole mode, and is mainly populated into self-attractive nonlinear region.

To our knowledge, the TBWs under self-attractive nonlinearity have not yet been investigated. Here we verify that they (the TBWs) exist but are always unstable under the constant self-focusing nonlinearity due to the presence of destructive coherence between adjacent gap modes, see a representative example of such solutions in Fig. \ref{fig2} (b) and its evolution in Fig. \ref{fig2} (d). Such instability can, however, be reined by the nonlinear lattice at or around $q=\pi$ (precise or almost complete commensurability between the nonlinear and linear lattices), see an example in Fig. \ref{fig2} (c) and the evolution in Fig. \ref{fig2} (e). This may be explained by the unique property of the nonlinear lattice with sign-alternating Kerr nonlinearity, which makes every gap mode stick to a single cell (strong focusing localization region) of the lattice and thus the destructive interaction between the adjacent modes can be greatly reduced. According to this argument, besides the six-peak localized states in Figs. \ref{fig2} (b) and  \ref{fig2} (c), one can construct stable TBWs with more peaks [see the inset in Fig. \ref{fig2} (a) for a fourteen-peak gap mode]. A noteworthy feature of the TBWs is that they are highly nonlinear, localized modes, which is a trait of fundamental gap solitons too, comparing the Figs. \ref{fig2} (b) and  \ref{fig2} (c) to the relevant subfigures in Fig. \ref{fig1}.

\begin{figure}[tbp]
\begin{center}
\includegraphics[width=1.0\columnwidth]{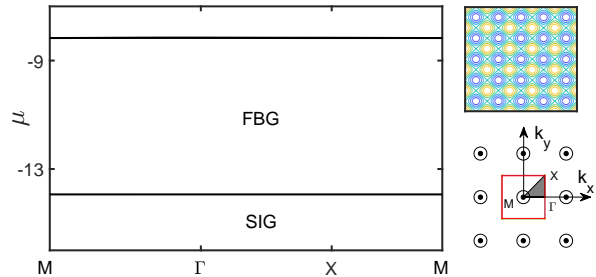}
\end{center}
\caption{ (Color online) Band structure for atomic Bloch waves in a 2D square lattice in the reduced Brillouin zone. Semi-infinite and first band gaps (SIG and FBG). Right top: contour plot of the linear lattice in Cartesian space. Right bottom: the first Brillouin zone (marked red) in the reciprocal lattice space; marked are the relevant high-symmetry points (M, $\Gamma$, X).}
\label{fig3}
\end{figure}

Recall that the 1D TBWs can only be stabilized by commensurate lattices
, thus, below we will focus our theoretical framework on such commensurate lattices, and compare them to that of constant self-focusing nonlinearity.

\emph{2D gap solitons and TBWs  under focusing segment of the nonlinearity}.---We now study the existence and stability of 2D gap modes under focusing (nonlinearity) segment of the nonlinear lattice, which is a challenging problem thanks to critical collapse is an intrinsic characteristic in 2D self-focusing media.

\begin{figure}[tbp]
\begin{center}
\includegraphics[width=1.0\columnwidth]{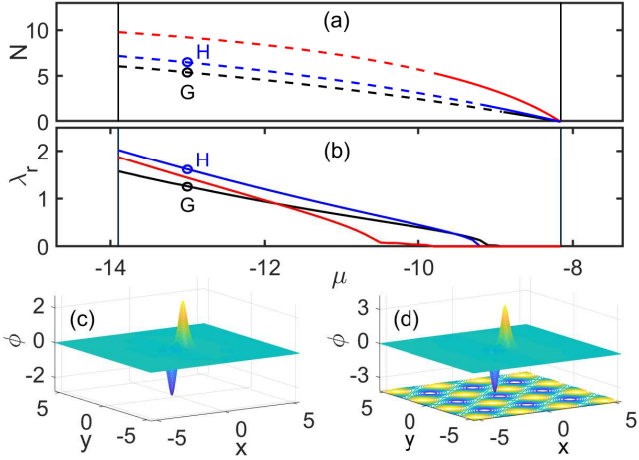}
\end{center}
\caption{(Color online)  (a) Norm $N$ versus $\mu$ for family of 2D fundamental gap solitons at attractive nonlinearity, under constant nonlinearity ($g_1=-1$, $g_0=0$, black line), and the nonlinear lattice (Eq. (\ref{NL2d})) with equal (blue line) and unequal (red line) amplitudes. (b) The relevant largest real part of instability growth rate $\lambda_r$ versus $\mu$. Typical 3D profiles of the gap solitons at $\mu=-13$ marked by circles G and H are showed in (c) and (d). In (a) and the Fig. \ref{fig5}(a) below, continuous line denotes stable subfamily, while the dotted line refers to unstable segment. Contour plot in (d) shows structure of the nonlinear lattice, darker blue contours implies self-focusing regions. }
\label{fig4}
\end{figure}

Band spectrum for atomic Bloch waves in the square lattice ($V_L(x,y)=10 [\cos(2x)+\cos(2y)]$), shown as chemical potential $\mu$ (energy) of the Bloch states, in the first Brillouin zone is depicted in Fig. \ref{fig3}. The spectrum diagram is denoted within the reciprocal lattice space, and the band-gap structure of the spectrum is obtained along the high-symmetry points of the irreducible zone. Showing is only for the first band gap since it is the region we are particularly interested in.

Two kinds of 2D nonlinear lattices with alternating focusing and defocusing cubic terms are taken: the sign-changing nonlinear strength (Eq. (\ref{NL2d})) with equal ($g_1=0, g_0=-1/2$) and unequal ($g_1=2, g_0=-3/2$) amplitudes. 
In Fig. \ref{fig4}(a), the $N(\mu)$ curve shows the existence of 2D fundamental gap solitons under the action of constant focusing nonlinearity, it is observed that their stability domain is only within a narrow region near the upper edge of the bandgap, conforming to those reported in similar setting \cite{DGS-2D}. The display of Fig. \ref{fig4}(a) also includes cases of the two types of nonlinear lattices under consideration, shown that the introduced periodic modulated nonlinearity can extend the relevant stability areas for the fundamental gap solitons, as exemplified by the linear stability analysis in Fig. \ref{fig4}(b) too.

\begin{figure}[tbp]
\begin{center}
\includegraphics[width=1.0\columnwidth]{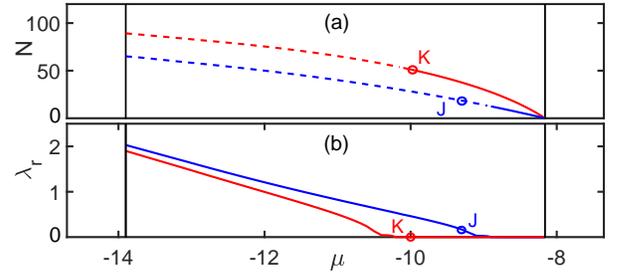}
\end{center}
\caption{ (Color online) (a) Norm $N$ versus $\mu$ for families of 2D TBWs composed of eighteen-peak localized states, supported by two kinds of periodic nonlinearities: the nonlinear lattice (Eq. (\ref{NL2d})) with equal (blue line) and unequal (red line) amplitudes. (b) The instability growth rate, $\lambda_r$, measuring the localized instability of 2D eighteen-peak TBWs. Circles J and K correspond to examples of unstable and stable eighteen-peak TBWs shown in Fig. \ref{fig6}.}
\label{fig5}
\end{figure}

While both of them can support 2D multiple-peak localized states (TBWs), the stable TBWs for the former is only restricted to a limited region ($\mu \in [-9, -8.3]$), the latter has a much wider stability region ($\mu \in [-10.2, -8.3]$), verified by linear-stability analysis and direct simulations, see Figs. \ref{fig5}(a) and \ref{fig5}(b). We thus conclude that the stability region of the 2D TBWs is much smaller than that of instability counterpart, owning to the growth rates (the corresponding real parts of the eigenvalues, $\lambda_r$) accounting for the instability decrease when the chemical potential $\mu$ is approaching the second Bloch band.  This type of the stability is specific to the 2D gap solitons and TBWs (under the action of self-focusing nonlinearity), which is in striking constrast to that of 1D scenarios in Figs. \ref{fig1} and \ref{fig2} where the stable gap modes can cover almost the whole first finite band gap.

\begin{figure}[tbp]
\begin{center}
\includegraphics[width=1.0\columnwidth]{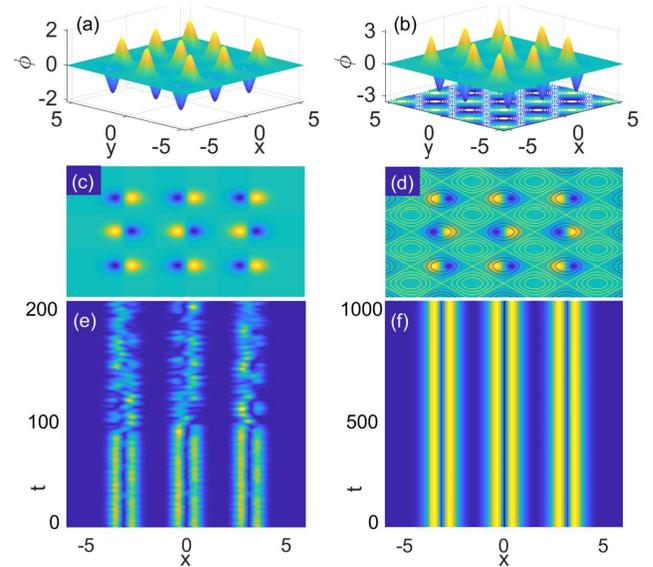}
\end{center}
\caption{Typical profiles of the unstable (a) and stable (b) 2D eighteen-peak ($6\times3$) localized states, and their contour plots in (c) and (d); depicted in (e) and (f) are the corresponding perturbed dynamics at $y=0$. Blue shading in (b) and (d) corresponds to self-focusing regions of the nonlinear lattice, canary yellow shading is for self-defocusing region.}
\label{fig6}
\end{figure}

Depicted in Figs. \ref{fig6}(a) and \ref{fig6}(b) are, respectively, typical stationary profiles of the unstable and stable 2D eighteen-peak localized states. Their contour plots and dynamics over time are shown in Figs. \ref{fig6}(c) and \ref{fig6}(d), and Figs. \ref{fig6}(e) and \ref{fig6}(f).
It is seen that an unstable TBW oscillate rapidly at first and then loses its shape after some time evolution, while for a stable one, its shape and amplitude still remain the same during dynamics evolution.

\emph{Experimental discussion}.---As mentioned above, the localized gap modes (gap solitons and TBWs) in media with constant defocusing nonlinearity have been realized both in optical \cite{GS-FBG, SS-AWA-2D, SGS-WA, GS-WA, CD-GDS-HPL, GS-NOIL}) and BECs experiments \cite{GS-SM1, GS-SM2, GS-SM3, GS-BEC}, the possibility to observe them under the action of self-focusing nonlinearity is a new question to be answered. The combined linear-nonlinear lattices model introduced here is applicable to nonlinear optics by filling the voids of photonic crystals with tailored nonlinear materials \cite{NL}, and to BECs by adopting the two powerful experimental skills---the optical lattice \cite{BEC-Rev1, BEC-Rev2} and Feshbach resonance controlled by spatially patterned fields \cite{nonuniform-Feshbach, Cce-FR, QD-BECs, Feshbach-magnetic, SVL-UA, FR-UG, FRM-BEC}; the predicted solutions supported by self-focusing nonlinearity can, therefore, be readily observed in experiments. The predicted TBWs supported by self-focusing nonlinearity (self-focusing segment of the nonlinear lattice) may also be observed experimentally in deep optical lattices where the quantum effects \cite{MBP-UCG} are nonnegligible, as suggested by the related work in [23].  For more specific experimental settings for realizing the nonlinear lattices introduced here, we refer the readers to consult the  review in \cite{NL} in detail. Note in passing that the first experimental observation of BECs gap solitons of atoms with repulsive atom-atom interaction (defocusing nonlinearity) is limited to low atom number \cite{GS-BEC}, the predicted gap solitons and TBWs will shed new light on the realization of localized gap modes in BECs for atoms with attractive interaction, and particularly, the stable TBWs upheld by self-focusing nonlinearity could be created in one and two dimensions and for an arbitrarily controllable number of atoms, making the observation more accessible in a general laboratory circumstance.

\emph{Conclusion}.---We studied the existence of two types of 1D and 2D self-trapped localized states, gap solitons and truncated nonlinear Bloch waves (TBWs), in an optical lattice and a nonlinear lattice (periodic variation of cubic Kerr nonlinear coefficient), under the focusing (attractive) nonlinearity. Properties and the formation mechanism of both localized states have been examined, and the relevant stability regions were collected by means of linear-stability analysis and identified via direct simulations. Theoretical descriptions of the spatially localized states of gap types so far boiled down to understanding the interplay of periodic potentials and focusing nonlinearity, this article predicts the first TBWs in the periodic physical systems but supported by focusing nonlinearity, thus opening up new possibilities of studying self-localization or self-trapping phenomena \cite{dipolar17,SOC18} in the same or similar models on the theoretical side, and most interestingly, of inspiring the experimentalists in various fields like the BECs and optics.

\section{Acknowledgment}

This work was supported, in part, by the NSFC, China (project Nos. 61690224, 61690222), and by the Youth Innovation Promotion Association of the Chinese Academy of Sciences (project No. 2016357).

\end{document}